\documentclass{JHEP3} 
\usepackage{epsfig}
\def \be {\begin{equation}}
\def \ee {\end{equation}}
\def \bea {\begin{eqnarray}}
\def \eea {\end{eqnarray}}
\def \nn {\nonumber}
\def \ba {\begin{array}}
\def \ea {\end{array}}

\def \a {\alpha}

\def \p {\partial}

\title{Supergravity Null Scissors and Super-Crosses}

\author{Bin Chen\\
Physics Division\\
National Center for Theoretical Sciences\\
Hsinchu 300, Taiwan\\
\email{bchen@phys.cts.nthu.edu.tw} }

\author{Chiang-Mei Chen\\
Department of Physics\\
National Taiwan University\\
Taipei 106, Taiwan\\
\email{cmchen@phys.ntu.edu.tw} }

\author{Feng-Li Lin\\
Physics Division\\
National Center for Theoretical Sciences\\
Hsinchu 300, Taiwan\\
\email{fllin@phys.cts.nthu.edu.tw} }

\abstract{ In this paper we construct the supergravity solutions
for the orthogonally intersecting null scissors and the fluxed
D-strings. We name the latter as the super-crosses according to
their shape. It turns out that the smeared solutions are U-dual
related to the intersecting $(p,q)$-strings. Their open string
properties are also studied. As a by-product, we clarify the
supersymmetry conditions of D2-D2 pairs with most generic fluxes.}

\preprint{\ hep-th/0303156} 

\begin{document}

\section{Introduction}
   String theory in the cosmological background is a challenging
subject to explore. One essential issue is to understand the
spacelike singularity in the context of string theory. One could
expect that there should have no singularity  due to the limit on
the minimal length scale in string theory. Or in other words, the
singularities would be resolved. This is a well-known phenomenon
in the case of resolution of orbifold and conifold singularities.

 A toy solvable model for the above purpose is the so called
null brane background discovered by Simon \cite{Simon}. Many
studies on this model have been done \cite{nullbrane}, especially
on the issue of the singularity resolution. Unfortunately the
study in \cite{nullbrane} indicates that there is an instability
to form black hole near the orbifold singularity due to the large
blue shift effect. Further study on the nature of the singularity
is obscured by the non-linear nature of the closed string theory.

The blue shift effect of the null brane frustrates the original
plan of studying the resolution of the cosmic singularities by
string theory, however, the mathematical structure of the model is
by itself interesting. This motivates people to study the open
string analogue of the null brane, the so called null scissors as
proposed and studied in \cite{Bachas} and \cite{Myers} (see also
\cite{Okuyama} and \cite{CO}), and to understand the nature of the
singularity without facing the non-linearity associated with
gravity. Null scissor is the configuration of two moving
intersecting D-strings with the intersection moving at the speed
of light. Surprisingly, a classical instability of the null
scissor is identified as the failure for the charged string to
catch up the motion of the intersecting point even though the open
string theory preserves $1/4$ supersymmetry. However, authors in
\cite{Bachas} tend to believe that the singularity will be
resolved though the detailed dynamics is not completely clear yet.

   As pointed out in \cite{Myers} the null scissor configurations are
closely related to the D-brane objects with nonzero fluxes, which
preserves unexpected supersymmetry and can be understood as the
U-duals of the known supersymmetric intersecting branes. Of among
them, the supertube \cite{supertube,Townsend3,Ohta} is the most
important and inspiring one. The U-duals of the supergravity
solution of the supertube has been studied in the recent paper
\cite{LMM}.

   The supersymmetry of the null scissors is also coming out of
surprise as for the one of supertube.  However, there is yet no
supergravity solution for the null scissor and its U-duals. The
authors of \cite{LMM} commented that the boosted string, which is
one component of the null scissor, can be obtained by T-dualizing
their solution for $D2$-$D0$-$F1$ configuration. We will see that
the full solution of the null scissor will arise from the U-dual
of the well-known $D1\perp F1$ configuration.

   In this paper we will fulfill the supergravity construction of
the orthogonally intersecting null scissors and its U-duals. One
of its U-duals is the orthogonally intersecting fluxed D-strings,
which is named as the ``super-cross" for the sake of its shape.
The key point of the construction is the realization of the
boosted D-string as the double T-dual of the $(p,q)$-string. From
this fact we can relate the orthogonally intersecting null
scissors to the S-dual of the well-known D1$\perp$F1 configuration
\cite{Gauntlett}.

   This paper is organized as follows. In the next section we
analyze the supersymmetry conditions of the D2-D2 pairs with the
most general E and B fluxes on them. The technical details of
checking the open string partition functions are also given in the
Appendix A. We then T-dualize the supersymmetry conditions to the
ones in IIB string for the null scissors and super-crosses. In
section 3, we construct the supergravity solutions of the null
scissors and super-crosses. The distinction between the KK
D-string and the boosted D-string is given in 3.1, the
supergravity solutions of the super-crosses are constructed from
the S-dual of D1$\perp$F1 in 3.2 and are shown to be double T-dual
to the orthogonally intersecting null scissors. The double
T-duality rule is given in the Appendix B. The non-existence of
the decoupling limit of the solutions is commented in 3.3.
Generalization to the wiggled solutions is discussed in 3.4. In
the section 4 we discuss the classical stability issue of the
super-cross. Finally we conclude our paper in section 5.

\section{Supersymmetric fluxed D2-D2 pairs and their T-duals}
  In this section we generalize the supersymmetry fluxed D2-brane pair
studied in \cite{Myers} to the more general case where the
electric fluxes on different D2-branes can point to different
directions. After T-duality, we can arrive not just the null
scissors but also the one with fluxes on D-strings so that the
speed of the intersecting point can be subliminal.

   Following \cite{Myers}, the supersymmetry constraints for
the flat D2-D2 pair are
\bea
\label{susy1}
&&\Gamma^{(1)} \epsilon= \epsilon\;, \\
\label{susy2} && \Gamma^{(2)} \epsilon=\epsilon\;,\\
\label{susy3}&& [\Gamma^{(1)}\;, \Gamma^{(2)}]\epsilon=0\;, \eea
where $\epsilon$ is the 10-dim. Killing spinor and
$\Gamma^{(i)}$'s are the Hermitian traceless product structure
given by the $\kappa$-symmetry \cite{Bergshoeff}, and takes the
following form in the static gauge \be \Gamma^{(i)} ={1\over {\cal
L}_{i}} [\Gamma_{012}+(\Gamma_2 F^{(i)}_{01}+\Gamma_0
F^{(i)}_{12}+\Gamma_1 F^{(i)}_{20})\Gamma_{11}]\;. \ee
 Here
$\Gamma_{m}$'s, with $m=0\dots 9$, are the constant 10-dim. Dirac
matrices such that $\{\Gamma_{m},\Gamma_{n}\}=2\eta_{mn}$, and
$\Gamma_{11}=\Gamma_0 \cdots \Gamma_9$. Also, $\Gamma_{m\dots
k}\equiv \Gamma_{m}\cdots \Gamma_{k}$. The ${\cal L}_i$ is the DBI
lagrangian for each D2. For simplicity, we have set the constant
$2\pi \ell^2_s$ in front of the generalized Born-Infeld field $F$
to unity.

   The difference of our analysis from the one in \cite{Myers} is
that we will turn on the generic background field
\be
F^{(i)}=E_i dx^2 \wedge dx^0+ G_i dx^1 \wedge dx^0+ B_i dx^2
\wedge dx^1\;.
\ee This gives
\be
{\cal L}_{i}=\sqrt{1-(E_i^2+G_i^2-B_i^2)}\;.
\label{DBI}
\ee

  From (\ref{susy1}) to (\ref{susy3}) we can turn the compatibility
condition (\ref{susy3}) into the the additional traceless
Hermitian product structure \be
 \Gamma^\prime=\frac{(E_1G_2-E_2G_1)\Gamma_0+(E_1B_2-E_2B_1)\Gamma_2+(G_1B_2-G_2B_1)\Gamma_1}{{\cal
 L}_2-{\cal L}_1}
\ee
 satisfying
  \be \Gamma^\prime \epsilon=\epsilon \label{susy4} \ee and \be
[\Gamma^\prime, \Gamma^{(i)}]=0 \ee so that the new supersymmetry
conditions (\ref{susy1}), (\ref{susy2}) and (\ref{susy4}) could be
diagonalized simultaneously. From (\ref{susy4}) we shall require
\be
 \Gamma^{\prime 2}=1,
\ee
which gives
\bea
 G^2_2&=&\frac{-(E_1B_2-B_1E_2)^2-(E_1\mp E_2)^2+(B_1 \mp B_2)^2}{{\cal
 L}^2_1} \label{g21}
\eea
if we set $G_1=0$ by using the freedom of rotational symmetry.

    As known the above condition is only necessary but may not
 be sufficient. We can pin down the relative sign by setting $G_2=0$ and compare it with
 the results in \cite{Myers}, then we
 arrive
 \be
 \Gamma^\prime\epsilon=\pm \epsilon, \hspace{5ex}
 G_2=\pm\sqrt{\frac{-(E_1B_2-B_1E_2)^2-(E_1-E_2)^2+(B_1-B_2)^2}{{\cal
 L}^2_1}}\;. \label{susydd}
 \ee

 Moreover, we need to impose the following relation
 \bea
 E_1E_2-B_1B_2 -1\leq 0 \label{DD}
 \eea
 to ensure the consistent D2-D2 supersymmetry condition.

Similarly we can consider the supersymmetry conditions for the
D2-antiD2 with various gauge fluxes. Now, the supersymmetry
conditions are then
\be
 \Gamma^{(1)}\epsilon=\epsilon, \hspace{5ex} \Gamma^{(2)}\epsilon=-\epsilon
\ee
 and $[\Gamma^{(1)}, \Gamma^{(2)}]\epsilon=0$. The corresponding additional
 Hermitian product structure to the compatibility condition is
 \be
 \tilde{\Gamma}= -\frac{(E_1G_2-E_2G_1)\Gamma_0+(E_1B_2-E_2B_1)\Gamma_2+(G_1B_2-G_2B_1)\Gamma_1}{{\cal
 L}_2+{\cal L}_1}.
 \ee
 The condition $ \tilde{\Gamma}^2=1$ will yield the same supersymmetry
condition (\ref{susydd}), i.e. with $\Gamma^{\prime}$ replaced by
$\tilde{\Gamma}$, but with a different consistency condition
 \bea
 E_1E_2-B_1B_2 -1\geq 0 \label{DDbar}\;.
\eea

As a consistent check of these supersymmetry conditions, one could
calculate the 1-loop partition function of open string between
D2s. In Appendix A, using lightcone boundary state formalism
\cite{Green}, we show that the partition functions vanish in both
the case of D2D2 and D2-antiD2 under corresponding supersymmetry
conditions.

  Before ending this section we will look into the T-dual D-string
configurations \footnote{In the appendix of \cite{Myers} one can
find the T-dual transformation between a single fluxed D2 and a
single fluxed moving D-string at an angle.}. Our main interest is
focused on the case $G_1=0$.
  If we perform the T-duality along the $x^1$ direction, then
we will get two D-strings spanning an angle $\theta$, with one of
them being static and carrying E-flux $e_1$, and the other being
moving with the normal speed $\beta_2$ and carrying E-flux $e_2$,
the supersymmetry condition (\ref{susydd}) then becomes \be
\label{ddsusy1}
-\beta_2^2(1-e_1^2)+\sin^2\theta=e_1^2+e_2^2(1-\beta_2^2)-2e_1
e_2\sqrt{1-\beta_2^2} \cos\theta\;. \ee

  However, if we perform the T-duality along the $x^2$
direction, we will instead obtain two moving D-strings spanning an
angle $\theta$, with only one of them carrying flux $e_2$, and the
supersymmetry condition (\ref{susydd}) becomes \be \label{ddsusy2}
-e_2^2(1-\beta_1^2)(1-\beta_2^2)+\sin^2\theta=\beta_1^2+\beta_2^2-2\beta_1
\beta_2\cos\theta\;. \ee

  One of the simple configurations from (\ref{ddsusy1}) is to set
$\beta_2=0$, and  we get the static fluxed scissors constrained by
\be
\label{flux}
e_1^2+e_2^2-2e_1 e_2 \cos\theta=\sin^2\theta\;.
\ee
The {\it super-crosses} are the ones with $\theta=\pi/2$, which
will be the starting point to obtain the supergravity description
in the next section.

  On the other hand, by setting $e_2=0$ of (\ref{ddsusy2}) we
get the moving null scissors with the intersecting point moving at
the speed of light, which can be seen by re-writing the
supersymmetry condition into
\be
\label{speed}
(\beta_1 \csc\theta + \beta_2 \cot\theta)^2+ \beta_2^2=1\;.
\ee

   From the similarity of the form (\ref{flux}) and (\ref{speed}), it
is clear that we can transmute super null scissors and
super-crosses by double T-duality along the orthogonal directions.

\section{Supergravity Solutions}
In this section we would like to construct the supergravity
solutions for the null scissors and super-crosses by using various
T and S dualities. Moreover, they can be related to each other by
the S-duality and the doubled T-duality, namely, T-dualizing twice
along two orthogonal directions. We summarize the relations in a
duality map. In the appendix B, we list the formulae for the
double T-duality.

\subsection{From D1${\perp}$F1 and $(p,q)$-string to KK and boosted D-strings}

Before going to null scissors and super-crosses, we will start
with the simpler configurations, namely, the D1${\perp}$F1 and
(p,q)-string; both are supersymmetric configurations. After the
double T-duality of both configurations, we will get the D-string
with Kalzua-Klein(KK) momentum and the boosted D-string
respectively.

It is known that a fundamental string orthogonally intersecting
with Dp-brane preserves $1/4$ supersymmetry \cite{Gauntlett}.
Moreover, the supergravity solution can be written down explicitly
according to the intersecting rules for the harmonic functions.
For $p=1$ with D-string lying on $x$-axis and F-string on
$y$-axis, which will be our starting point for constructing the
other solutions, the configuration is
\begin{eqnarray}
\label{D1F1}
ds^2 &=& e^{\frac{\phi}2} \left[ - H_f^{-\frac34} H_d^{-\frac34}
dt^2 + H_f^{\frac14} H_d^{-\frac34} dx^2 + H_f^{-\frac34}
H_d^{\frac14} dy^2 + H_f ^{\frac14} H_d^{\frac14} ( dr^2 + r^2
d\Omega^2_6) \right],
\nn\\
e^{2\phi} &=& H_f^{-1} H_d, \qquad B_{[2]} = - (H_f^{-1} - 1) dt
\wedge dy, \qquad C_{[2]} = - (H_d^{-1} - 1) dt \wedge dx,
\end{eqnarray}
where the harmonic functions are given by $H_f(r) = 1 + Q_f / r^5$
and $H_d(r) = 1 + Q_d / r^5$ with parameters $Q_f, Q_d$
characterizing the charges of F-string and D-string.

   After double T-duality\footnote{The order of these two T-dualities is immaterial.}
as given in the Appendix B, which changes the location of D-string
from $x$-axis to $y$-axis and converts the F1 charge into a KK
moment on the $y$ direction, we get the D-string with KK momentum
as expected, the solution is
\begin{eqnarray}\label{KKD}
ds^2 &=& H_d^{-\frac12} \left[ - dt^2 + (H_f - 1) (dt + dy)^2 +
dy^2 \right] + H_d^{\frac12} (dx^2 + dr^2 + r^2 d\Omega^2_6)\;,
\nonumber
\\
&=& H_d^{-\frac12} \left[ - du dv + (H_f - 1) du^2 \right] +
H_d^{\frac12} (dx^2 + dr^2 + r^2 d\Omega^2_6)\;,
\nn\\
e^{2\phi} &=& H_d, \qquad C_{[2]} = - (H_d^{-1} - 1) dt \wedge
dy\;,
\end{eqnarray}
where the light cone coordinates are defined as $u = t + y$ and
$v= t - y$. For simplicity, we will call this configuration KK
D-string.

   On the other hand, if we start with a $(p,q)$-string located along
the $x$-axis and smeared along the $y$-axis,
\begin{eqnarray}
\label{pqstring}
ds^2 &=& e^{\frac{\phi}2} \left[ H^{-\frac34} ( - dt^2 + dx^2 ) +
H^{\frac14} ( dy^2 + dr^2 + r^2 d\Omega^2_6 ) \right]\;,\nn
\\
e^{2\phi} &=& \frac{(p^2 + q^2 H)^2}{(p^2 + q^2)^2 H}, \qquad
C_{[0]} = - \frac{p q(1 - H)}{p^2 + q^2 H}\;,\nn
\\
B_{[2]} &=& - p (H^{-1} - 1) dt \wedge dx, \qquad C_{[2]} = -q
(H^{-1} - 1) dt \wedge dx\;,
\end{eqnarray}
where the harmonic function $H = 1 + Q / r^5$. This solution can
be obtained by applying S-duality to either the F-string or
D-string solution.

  Performing the double T-duality of this configuration,
we obtain a D-string boosted along the transverse direction, and
the solution looks as
\begin{eqnarray}\label{boostD}
ds^2 &=& e^{-\phi} \left\{ - e^{2\phi} H^{-1} dt^2  + H [ p(H^{-1}
- 1) dt - dx ]^2 + dy^2 \right\} + e^\phi (dr^2 + r^2
d\Omega^2_6),
\nn\\
e^{2\phi} &=& \frac{p^2 + q^2 H}{p^2 + q^2}, \qquad C_{[2]} = - q
e^{-2\phi} (1 - H) ( dt + \frac{p}{p^2 + q^2} dx) \wedge dy.
\end{eqnarray}
The location of D-string and the direction of the boost are
clearly shown in the expressions of potential and metric. To
distinguish from the previous configuration, we will call this the
boosted D-string. Similarly, we can construct the boosted F-string
configuration via S-duality.

   As an interesting exercise, we can get the $(p,q)$ string
configuration as the T-dual of $D2$-$D0$-$F1$ which is given in
the (5.8) of \cite{LMM} by T-dualizing the boosted D-string. To be
explicit, the T-dual of (5.8) in \cite{LMM} is
\bea
\label{LMMT}
ds^2&=&-f^{1\over2}g^{-1}dt^2+f^{1\over2}[dx^2+dy^2-(g^{-1}-1)(\cos\theta
dx+\sin\theta dy)^2+dr^2+r^2d\Omega^2_6]\;,\nn
\\
e^{2\phi}&=&f^2g^{-1}\;, \qquad C_{[0]}=(f^{-1}-1)\sinh\a
\\
B_{[2]}&=&(g^{-1}-1)\tanh\a(\cos\theta dx+\sin\theta dt)\wedge
dy=-\sinh\a C_{[2]}\;,\nn
\eea
where
\be
f=1+{Q'\over r^5}\;, \qquad g=1+{Q'\cosh^2\a \over r^5}\;.
\ee

  If $\theta=0$ this configuration is reduced to the
$(p,q)$-string configuration (\ref{pqstring}) if we identify
\be
p=\tanh\a\;, \qquad q=-{1\over \cosh\a}\;, \qquad Q=Q'\cosh^2\a\;.
\ee
Note that $p^2+q^2=1$ satisfies the charge quantization condition
of the $(p,q)$-string.

\subsection{Super crosses and null scissors}
  From the discussions in the previous subsection, it is now clear
that if we would like to have the null scissor configuration which
is the superposition of two boosted D-strings, we need to start
with the configuration of two orthogonally $(p,q)$-string, and
then perform the double T-duality.

The general supersymmetric configuration for the intersecting
$(p,q)$-string at any angle is not known, however, the one with
orthogonal intersection can be obtained by S-dualizing the
D1${\perp}$F1 configuration (\ref{D1F1}), and we obtain the
super-cross configuration as follows:
\begin{eqnarray}\label{SuperCross}
ds^2 &=& e^{\frac{\phi}2} \left[ - H_f^{-\frac34} H_d^{-\frac34}
dt^2 + H_f^{\frac14} H_d^{-\frac34} dx^2 + H_f^{-\frac34}
H_d^{\frac14} dy^2 + H_f ^{\frac14} H_d^{\frac14} ( dr^2 + r^2
d\Omega^2_6) \right],
\nn\\
e^{2\phi} &=& \frac{(p_1^2 H_d + p_2^2 H_f)^2}{(p_1^2 + p_2^2)^2
H_f H_d}, \qquad C_{[0]} = \frac{p_1 p_2(H_d - H_f)}{p_1^2 H_d +
p_2^2 H_f},
\nn\\
B_{[2]} &=& - dt \wedge \left[ p_2 (H_d^{-1} - 1) dx + p_1
(H_f^{-1} - 1) dy \right],
\nn\\
C_{[2]} &=& - dt \wedge \left[ p_1 (H_d^{-1} - 1) dx - p_2
(H_f^{-1} - 1) dy \right].
\end{eqnarray}
Note that the harmonic functions $H_{f,d}$ now becomes $1+{\a_p
Q_{f,d} \over r^5}$ with $\a_p=\sqrt{p_1^2+p_2^2}$ to have correct
$(p,q)$-string bound state charges.

  This solution describes the smeared configuration of two
orthogonally intersecting $(p,q)$-strings, namely $(p1,p2) \perp
(-p2,p1)$. However, interestingly it can also be considered as a
composed configuration of two superposed D1$\bot$F1 bound states,
namely $p_1$(D1$\bot$F1) $\oplus$ $p_2$(F1$\bot$D1). Since both
configurations carry the same charges we are not able to
distinguish them. The evidence for these twofold descriptions can
be explored by considering the following special limits: The
super-cross reduces to D1$\bot$F1 by simply setting either
$p_1=1,p_2=0$ or $p_1=0,p_2=1$ and to the $(p,q)$-strings by
assuming the $H_f=1,H_d=H,p_2=p,p_1=q$. This degeneracy is
understood as a coincidence due to the initial symmetric
configuration (\ref{D1F1}).

  Now we can obtain the orthogonally intersecting null scissor
configuration by performing the double T-duality on
(\ref{SuperCross}), we then get
\begin{eqnarray}\label{NullScissor}
ds^2 &=& e^{-\phi} \left\{ - e^{2\phi} H_f^{-1} H_d^{-1} dt^2 +
H_d [p_2(H_d^{-1} - 1) dt - dx]^2 + H_f [p_1(H_f^{-1} -1) dt -
dy]^2 \right\} \nonumber
\\
&+& e^\phi (dr^2 + r^2 d\Omega^2_6 ),
\nn\\
e^{2\phi} &=& \frac{p_1^2 H_d + p_2^2 H_f}{p_1^2 + p_2^2},
\\\nn
C_{[2]} &=& e^{-2\phi} \left\{ - dt \wedge \left[ p_2(1 - H_f) dx
+ p_1 (1 - H_d) dy \right] + \frac{p_1 p_2}{p_1^2 + p_2^2} (H_d -
H_f) dx \wedge dy \right\}\;.
\end{eqnarray}
In comparison with the single boosted D-string configuration
(\ref{boostD}) the null scissor configuration looks much like the
``superposition" of the two orthogonally intersecting boosted
D-strings as one should expect. Moreover, due to the degeneracy
between $(p1,p2) \perp (-p2,p1)$ and $p_1$(D1$\bot$F1) $\oplus$
$p_2$(F1$\bot$D1) the configuration (\ref{NullScissor}) should
also describe two orthogonally intersecting KK D-strings.

   As shown in section 2, the open string analysis suggested that
the E-fluxes on the super-cross should satisfy the relation
$e_1^2+e_2^2=1$ for orthogonal intersection. From the NS-NS 2-form
$B_{[2]}$ of configuration (\ref{SuperCross}) we find that the
E-fluxes per $(p,q)$-string bound state charge is just $p_1$ and
$p_2$  respectively, so that the supersymmetric condition requires
\be
\label{cq} p_1^2+p_2^2=1\;.
\ee
This condition is the same as the one imposing on the parameters
of the S-duality to recover the charge quantization, namely, the
$SO(2)$ subgroup of $SL(2,R)$ is selected \cite{sdual}. On the
other hand, if the supersymmetry condition is for the E-fluxes per
D-string charge, which is either $-p_1/p_2$ or $p_2/p_1$, it will
be in conflict with the charge quantization condition (\ref{cq}).

For the case $p_1^2+p_2^2=1$, in which the parameters can be
chosen as $p_1=\cos\omega, p_2=\sin\omega$, the solution has a
very simple form
\begin{eqnarray}
\label{Null}
ds^2 &=& e^{-\phi} \left[ - dt^2 + (H_d - 1) (\sin\omega dt +
dx)^2 + (H_f - 1) (\cos\omega dt + dy)^2 + dx^2 + dy^2 \right]
\nonumber
\\\nn
&+& e^\phi (dr^2 + r^2 d\Omega^2_6 ),
\\
e^{2\phi} &=& H_d \cos^2\omega + H_f \sin^2\omega,
\\\nn
C_{[2]} &=& e^{-2\phi} \{ - dt \wedge [\sin\omega (1\!-\!H_f) dx +
\cos\omega (1\!-\!H_d) dy] + \sin\omega \cos\omega (H_d\!-\!H_f)
dx \wedge dy \}.
\end{eqnarray}

Furthermore, for the more special case $H_f = H_d = f$, the above
configuration is reduced to a more suggestive form for null
scissor
\begin{eqnarray}
ds^2 &=& f^{-\frac12} \left\{ - dt^2 + (f-1) \left[ (\sin\omega dt
+ dx)^2 + (\cos\omega dt + dy)^2 \right] + dx^2 + dy^2 \right\}
\nonumber
\\
&+& f^{\frac12} (dr^2 + r^2 d\Omega_6^2),
\\\nn
e^{2\phi} &=& f, \qquad C_{[2]} = - (f^{-1} - 1) dt \wedge
(\sin\omega dx + \cos\omega dy).
\end{eqnarray}
In the appropriate gauge choice of potentials, one can easily see
that there are momenta both along $x$- and $y$-axes respectively
which can not be ``mingled'' by a rotation.

  Up to now we learn that different fluxed or boosted configurations
are related by the T-duality, S-duality or particular reductions,
we summarize these relations in a duality map in the Figure 1.

\FIGURE[ht]{ \epsfig{figure=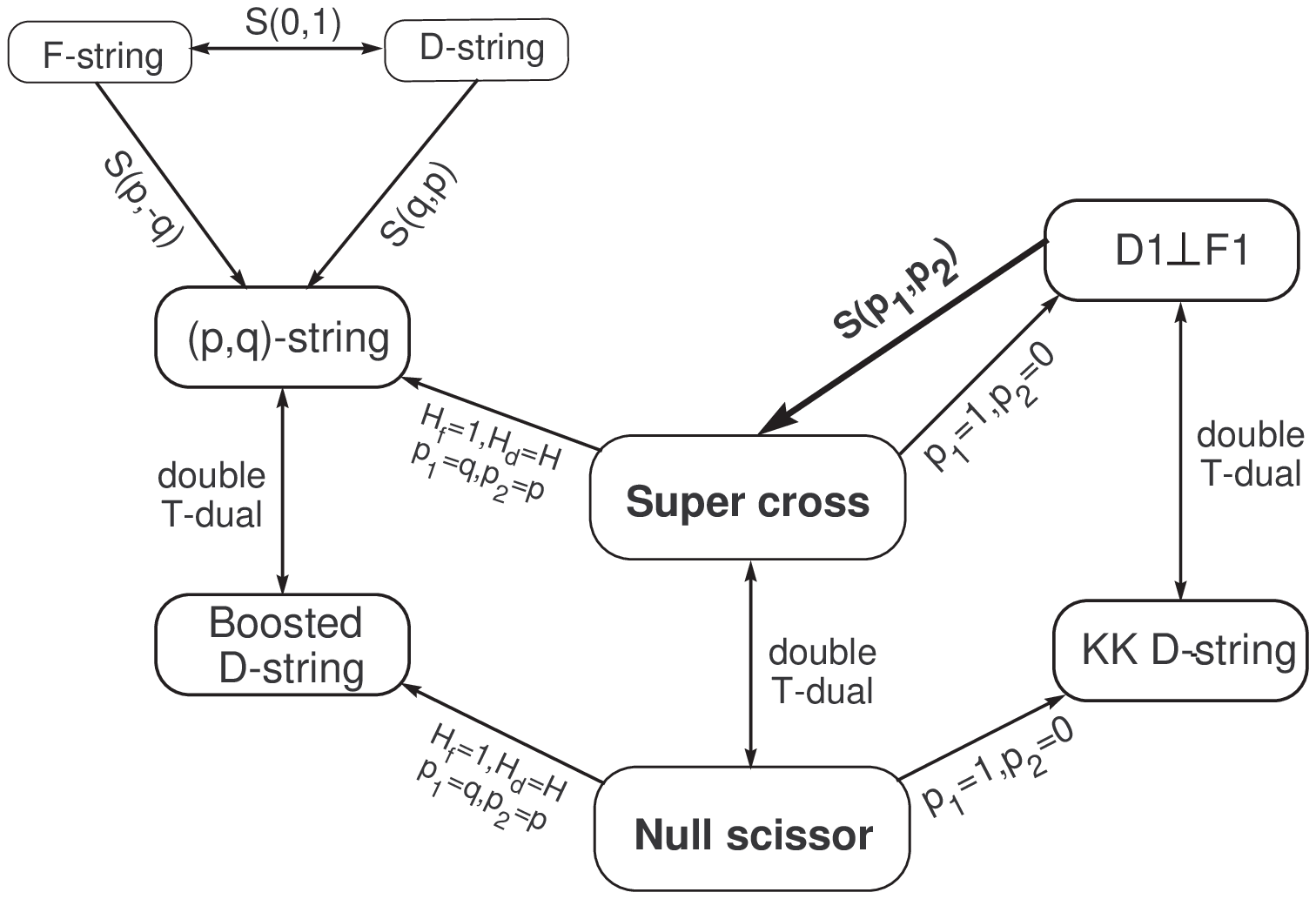,width=12cm}
 \caption{Duality Map}
 \label{fig}
}

  All the configurations in the above duality map are only for the
ones related to the orthogonally intersecting branes, it is
interesting to construct the general supergravity solution for the
null scissors with an angle less than $\pi/2$. It is easy to see
that these configurations cannot be obtained from the D1$\perp$F1
by the above S-duality and double T-duality. Even the double
T-duality along the different $x$-$y$ axes is just introduce a
trivial rotation on the S-duality parameters, we think that one
should start with the non-orthogonally intersecting supersymmetric
D1/F1 configuration in order to get the generic null scissors.  We
leave this construction for the future interest.

\subsection{Decoupling limit?}

  After getting the supergravity solutions for the null scissors
and the super-crosses, we would like to see that if there is any
dual description of Super Yang-Mills theory by taking some
decoupling limit. If yes, it could be possible for us to
understand the semi-classical instability issue of the null
scissors \cite{Bachas,Myers} from the dual gravity side.

   Unfortunately, it is easy to see from the solutions
(\ref{Null}) or (\ref{SuperCross}) that there exists no such
limit. The reason is as following: In order to have a sensible
decoupling limit as $\a' \to 0$, it  requires that $e^{\phi}$
scale as the $(\a')^0$ and the metric scale as $\a'$, however,
from the metric, we found that it is not possible to choose the
scaling behavior for $H_f$, $H_d$ and $r$ and to keep $g_{YM}$
fixed so that the metric will scale as $\a'$. In fact, this is
already the case for the configuration of D1$\perp$F1.

   On the other hand, since there is a possibility to have
critical E-field for the super-crosses, one may wonder if there
exits a NCOS limit \cite{NCOS}. However, the charged string
hanging between the un-fluxed and the fluxed D-strings makes the
critical field limit problematic since the two D-strings are not
parallel so that the charged string can not be polarized to have
zero tension. Moreover, as mentioned, our supergravity solution
goes back to the D1$\perp$F1 as we takes $p_1=0,1$, and we cannot
obtain the critical E-flux case from the solution.

\subsection{Wiggled super-crosses and null scissors}
The super-crosses and null scissors can be straightforwardly
generalized to a wiggled version. An easy way is starting from an
{\it oscillating string} \cite{CMP,LMM}, a generalization of the
KK string which has a transverse profile described by an arbitrary
function $\mathbf{F}(u)$,
\begin{eqnarray}\label{OscF1}
ds^2 &=& H_1^{-1}(\mathbf{z},u) \left[ - du dv + K(\mathbf{z},u)
du^2 + 2 A_i(\mathbf{z},u) dz^i du \right] + dx^2 + d\mathbf{z}^2,
\\
e^{2\phi} &=& H_1^{-1}(\mathbf{z},u),
\\
B_{[2]} &=& \frac12 \left( H_1^{-1}(\mathbf{z},u) - 1 \right) du
\wedge dv + H_1^{-1}(\mathbf{z},u) A_i(\mathbf{z},u) \, du \wedge
dz^i,
\end{eqnarray}
where the functions are defined as
\begin{eqnarray}
H_1(\mathbf{z},u) &=& 1 +
\frac{Q_1}{|\mathbf{z}-\mathbf{F}(u)|^5}, \qquad
H_2(\mathbf{z},u) = 1 + \frac{Q_2}{|\mathbf{z}-\mathbf{F}(u)|^5},
\\
K(\mathbf{z},u) &=& (H_1 - 1) (\partial_u \mathbf{F})^2 + (H_2 -
1), \qquad  A_i(\mathbf{z},u) = - (1 - H_1) \partial_u F_i.
\end{eqnarray}
Then all interesting solutions can be constructed at every step
of the following approaches.
\begin{eqnarray}
\mbox{Oscillating F-string} \quad
\stackrel{S(0,1)}{\longrightarrow}& \quad \mbox{Oscillating
D-string} \quad &\stackrel{T^2}{\longrightarrow} \quad
\mbox{Wiggled D1$\bot$F1} \nonumber
\\
\quad \stackrel{S(p_1,p_2)}{\longrightarrow}& \quad \mbox{Wiggled
Super crosses} \quad &\stackrel{T^2}{\longrightarrow} \quad
\mbox{Wiggled Null Scissors} \nonumber
\end{eqnarray}
Therefore, the wiggled D1$\bot$F1 solution is
\begin{eqnarray}
ds^2 &=& H'{}_2^{-1} H_1^{-\frac12} \left[ -(dt - A_i dz^i)^2 +
H'{}_2 dx^2 + H_1 dy^2 \right] + H_1^{\frac12} d\mathbf{z}^2 ,
\\
e^{2\phi} &=& H'{}_2^{-1} H_1, \qquad B_{[2]} = - (H'{}_2^{-1} -
1) dt \wedge dy - H'{}_2^{-1} A_i dy \wedge dz^i,
\\
C_{[2]} &=& - (H_1^{-1} - 1) dt \wedge dx + H_1^{-1} A_i dx
\wedge dz^i,
\end{eqnarray}
where $H'_2 = K + 1$.

   From this configuration one can perform the S-duality and the
double T-dualities as done in the previous subsection to obtain
the wiggled super-crosses and null scissors, the procedures are
straightforward and the final tedious results are omitted.

\section{Stability Analysis of Static Fluxed D-String Pair}

In this section, we will use the classical Super Yang-Mills theory
(SYM) to study the motion of the charged string hanging between
the two D-strings.

Start with the equation of motion for the adjoint scalar in the
$(1+1)$ U(2) SYM, that is
\be
D_aD^a
\Phi^{(i)}+{1\over2}[\Phi^{(j)},[\Phi^{(i)},\Phi^{(j)}]]=0\;,
\ee
where the $D^a=\p^a+i[A^a, \cdot]$ is the covariant derivative.

In order to understand the effect of the electric field on the
D-string to the motion of the charged string, we choose the
simplest supersymmetric background with the following vevs, \be
\label{vev} \Phi^{(2)}={V\over 2}(\tau_0+\tau_3)\;, \qquad A_1={et
\over 2}(\tau_0-\tau_3)\;, \ee where the static gauge is chosen so
that $A_0=0, \Phi^0=t, \Phi^1=x$, and \be V\equiv x\tan\theta
+{\beta t\over \cos\theta}\;. \ee Furthermore, the supersymmetry
condition requires \be e=\sqrt{\sin^2\theta -\beta^2 \over
1-\beta^2}\;. \ee As mentioned, this configuration describes a
pair of D-strings, one lies on the $x$-axis, and the other of
nonzero constant E-flux moves with speed $\beta$ in the direction
with an angle relative to $x$-axis. This configuration solves the
equation of motion but break the gauge symmetry to $U(1)\times
U(1)$.

  We introduce the transverse perturbation represented by a charged
string in the background of (\ref{vev}) as
\bea
\delta \Phi^{(\bot)}=\left(\ba{cc}
 0& \phi_+\\
 \phi_- &0
  \ea \right)\;,
\eea
and the equation of motion for the charged string perturbation is
reduced to
\be
\{ -\p_t^2+ \p_x^2\mp 2iet \p_x -(et)^2 -{1\over2}V^2
\}\phi_{\pm}=0\;,
\ee
or to a more physically transparent form as
\be
\label{stable}
\{ p_t^2-(p_x \pm A_x)^2-{1\over2}V^2 \}\phi_{\pm}=0\;,
\ee
where $p_t=i\p_t$, $p_x=-i\p_x$ and $A_x=F_{tx}t=et$. This
describes a pair of oppositely charged particles subjected to a
constant electric field and a simple-harmonic-like potential.

   If $\beta=0$, the configuration reduces to the static fluxed
scissors with $e_1=\sin\theta$ and $e_2=0$ of (\ref{flux}). For
$\theta=\pi/2$ this is just the special case of the super-cross.
In these cases, (\ref{stable}) just describes a relativistic
simple harmonic oscillator subjected to the constant electric
field.  As known in the non-relativistic case, this configuration
is localized and thus it is stable. The stability of the
relativistic case should remain if the quantum Schwinger effect of
pair production by the E-field is neglected. On the other hand, if
$\beta\ne 0$ the potential $V$ is both space and time dependent
and (\ref{stable}) is hard to solve, it is then not easy to tell
the stability in lack of the explicit solution.

 We then conclude that the static fluxed scissors should be
classically stable. This raises some issue of concern about the
stability in the T-dual related configurations. As we found in
this paper, the super-cross and null scissor are related by double
T-duality and both are related to the fluxed D2-brane pair by
T-duality.  However, the fluxed D2-brane and the super-cross are
classically stable but the null scissor is not \cite{Bachas,Myers}
although all are supersymmetric and have zero 1-loop result. These
facts indicate that the T-duality and the classical stability are
not commutative. Though that, we expect that quantum effect will
comes into play as the total energy goes high so that the net
effect will yield a stable configuration as indicated by the
1-loop result.

\section{Conclusions}

   In this paper we have constructed the supergravity
configurations for the orthogonally intersecting null scissors and
super-crosses. Also we have found a duality map between different
intersecting string configurations. As emphasized in the
introduction the key point in the construction is the realization
of the boosted D-string as the double T-dual of the
$(p,q)$-string. The more general supergravity null scissor
configurations with arbitrary intersecting angle are expected to
be constructed along the same line although one needs to first
obtain the more general starting configuration other than the
D1${\perp}$F1.

  One of our original motivation for the supergravity solution is
to understand the stability issue of the null scissors from the
dual gravity side. Unfortunately there exists no sensible
decoupling limit for these supergravity configurations. It is not
clear at all why such obstacle exist for the dual description.

  One can also follow the open string spectrum analysis for the
null scissor as done in \cite{Bachas} and do different T-duality
on the fluxed D2-brane pair to obtain the one for the super-cross
or static intersecting fluxed D-string. The resulting on-shell
condition is complicated and it is hard to extract any conclusive
result from that. However, we notice that in the analysis of
\cite{Bachas} the oscillator modes in one of the worldvolume
direction are suppressed so that the center of mass momentum of
the charged string is not dependent on the oscillator modes.  It
may be difficult to justify the suppression of these oscillator
modes as the total energy goes high.  This might be a loop hole to
the stability issue and the full picture of the resolution of the
open string singularity remains as a challenge.

\vskip 1cm

\section*{Acknowledgements}

B.C would like to thank the support of Prof. Kiwoon Choi on his
stay in KAIST, and also thank the hospitality of KIAS during his
visit.  The work of CMC is supported by National Science Council
and the CosPA project. FLL acknowledges the hospitality of the
Lorentz Center where this project was inspired.

\appendix

\section{Stability Analysis at 1-loop}

In this Appendix we will calculate the stringy 1-loop partition
function and verify the supersymmetry condition, i.e. zero
partition function, for the D2-D2 pair configuration discussed in
section 2.

Let's give a short review of the calculation in \cite{Townsend3},
where the light-cone gauge boundary state formalism has been
used\cite{Green}. In the light-cone boundary state formalism, the
Dp-brane could be taken as the `$(p+1)$' instanton with Euclidean
worldvolume which could be related to the ordinary Lorentz
worldvolume through double Wick rotation.

 The partition function between two parallel brane can be
 factorized
into
 \be
 Z=Z_0Z_{osci}
 \ee
 where
 \be
 Z_0={\cal L}_E(F^{(1)}){\cal L}_E(F^{(2)}) (Tr_vM_{rel}-Tr_sM_{rel})
 \ee
 Here, we have
 \be
 M_{rel}=M^T_2\cdot M_1
 \ee
 with $M$ being the SO(8) matrix
 \be
 M_{IJ}=\left(\ba{cc}
 M_{ij}& \\
  &I_{7-p}
  \ea \right)\;.
 \ee

For the generic flux configuration
 \be
 F=Edx^1\wedge dx^2+Gdx^3\wedge dx^1+Bdx^2 \wedge dx^3,
 \ee
 \bea
 M^v_{ij}&=& -(1-F)(1+F)^{-1} \nonumber\\
       &=& -\frac{1}{{\cal L}_E^2(F)}\left( \ba{ccc}
       1+B^2-E^2-G^2&2(BG-E)&2(EB+G)\\
       2(E+BG)&1+G^2-B^2-E^2&2(-B+EG)\\
       2(EB-G)&2(B+EG)&1+E^2-B^2-G^2
       \ea\right)\\
 \eea
 and
 \bea
 M^s_{ij}=\frac{1}{{\cal L}_E(F)}\gamma^{12\cdots
 8}(\gamma^{123}+E\gamma^3+B\gamma^1+G\gamma^2)
 \eea
 where ${\cal L}_E(F)=\sqrt{1+E^2+B^2+G^2}$.

In order to consider both D-D and D-antiD case, we take into
account of the rotation in the $(3,4)$ plane. Then the rotation
matrix becomes
 \be
 M=m^{-1}(\phi)\cdot M\cdot m(\phi)
 \ee
 where
 \be
 m^v(\phi)=\left(\ba{ccccc}
 1& & & & \\
  &1 & & &\\
  & & \cos(\phi)&-\sin(\phi)& \\
  & &\sin(\phi)&\cos(\phi) & \\
  & & & &I_4
  \ea \right)
  \ee
 in the vector representation and
 \be
 m^s(\phi)=\exp(\frac{\phi}{2}\gamma^{34})=\cos(\frac{\phi}{2})I+\sin(\frac{\phi}{2})\gamma^{34}
 \ee
 in the spinor representation.

 From these profiles, we can read out the zero-mode contribution to
the partition function:
 \be
 Z_0=\frac{\lambda}{{\cal L}_E(F_2){\cal L}_E(F_1)}-\rho
 \ee
 where\footnote{Here our interests are on the D-D and D-antiD brane
 configuration so we don't put $\cos{2\phi}$ into the relation
 since $\cos(2\phi)=1$ in both cases.}
 \bea
 \lambda&=&
 8[E_1E_2+B_1B_2G_1G_2+(E_1E_2B_1B_2+G_1G_2+E_1E_2G_1G_2+B_1B_2)\cos(\phi)]
 \nonumber \\
 &+&4(1+E^2_1E^2_2+B^2_1B^2_2+G^2_1G^2_2)+4{\cal L}_E^2(F_2){\cal L}_E^2(F_1)
 \eea
 and
 \bea
 \rho=8[(1+E_1E_2)\cos(\phi)+B_1B_2+G_1G_2]
 \eea

 To compare with the brane configuration we discussed in the last
 section, let us do double Wick rotation  and set $G_1=0$. Then we find that  the zero-mode
 contribution to the partition function to be
 \bea
 \lambda&=&8(-E_1E_2-E_1E_2B_1B_2\cos(\phi)+B_1B_2\cos(\phi))\nonumber\\
    &+&4(1+E^2_1E^2_2+B^2_1B^2_2)+4{\cal L}^2_1{\cal L}^2_2\\
 \rho&=&8((1-E_1E_2)\cos(\phi)+B_1B_2)
 \eea
where  ${\cal L}_{i}$'s are defined in (\ref{DBI}) with $G_1=0$.

Firstly, let us consider the D-D case where $\phi=0$ and put
(\ref{g21}) into the relation, then we obtain \be
 Z_0=8|1-E_1E_2+B_1B_2|-8(1-E_1E_2+B_1B_2)=0
\ee under the consistency condition (\ref{DD}). This confirms the
supersymmetry condition found in section 2. Similarly this is true
for the D-antiD case \footnote{One should  let $B_2 \rightarrow
-B_2$ because the ant-D2 is obtained by flipping D2 so that the
direction of $B2$ is reversed with respect to $B_1$.} which
corresponds to $\phi=\pi$.

\section{Formula for double T-dual super-cross}
Let us consider a general T-dual result of the super-cross,
starting from the following general type of solution
\begin{eqnarray}\label{GenForm}
ds^2 &=& G_{tt} dt^2 + G_{xx} dx^2 + G_{yy}dy^2 + 2 G_{xy} dx dy +
G_{rr}( dr^2 + r^2 d\Omega_6^2 ),
\nn\\
B_{[2]} &=& B_{tx} dt \wedge dx + B_{ty} dt \wedge dy, \qquad
C_{[2]} = C_{tx} dt \wedge dx + C_{ty} dt \wedge dy,
\end{eqnarray}
with non-vanishing axion $C_{[0]} = C_0$ and dilaton $\phi_0$.

After T-duality along x-axis, we obtain the new configuration as
following
\begin{eqnarray}
ds^2 &=& \left( G_{tt} + \frac{B^2_{tx}}{G_{xx}} \right) dt^2 +
\frac1{G_{xx}} dx^2 + \frac{G}{G_{xx}} dy^2 + 2
\frac{B_{tx}}{G_{xx}} dt dx + G_{rr} ( dr^2 + r^2 d \Omega_6^2 ),
\nn\\
e^{2\phi} &=& \frac{e^{2\phi_0}}{G_{xx}}, \qquad B_{[2]} = -
\frac{G_{xy}}{G_{xx}} dx \wedge dy + \frac{B}{G_{xx}} dt \wedge
dy,
\nn\\
C_{[1]} &=& ( C_{tx} + C_0 B_{tx} ) dt + C_0 dx, \qquad C_{[3]} =
\left( -C_{ty} + \frac{C_{tx} G_{xy}}{G_{xx}} \right) dt \wedge dx
\wedge dy,
\end{eqnarray}
where
\begin{equation}
G \equiv G_{xx} G_{yy} - G_{xy}^2, \qquad B \equiv B_{ty} G_{xx} -
B_{tx} G_{xy}\;.
\end{equation}

Then we take another T-duality on y-axis and arrive
\begin{eqnarray}
ds^2 &=& \left( G_{tt} + \frac{B' B_{tx} + B B_{ty}}{G} \right)
dt^2 + \frac{G_{yy}}{G} dx^2 + \frac{G_{xx}}{G} dy^2 \nonumber
\nn\\
&& + 2 \frac{B'}{G} dt dx + 2 \frac{B}{G} dt dy  + 2
\frac{G_{xy}}{G} dx dy + G_{rr} ( dr^2 + r^2 d\Omega_6^2 ),
\nn\\
e^{2\phi} &=& \frac{e^{2\phi_0}}{G}, \qquad B_{[2]} = 0
\nn\\
C_{[2]} &=& - (C_{ty} + C_0 B_{ty}) dt \wedge dx + (C_{tx} + C_0
B_{tx}) dt \wedge dy + C_0 dx \wedge dy,
\end{eqnarray}
where
\begin{equation}
B' \equiv B_{tx} G_{yy} - B_{ty} G_{xy}\;.
\end{equation}


\end{document}